# Radiofrequency response of the optically detected level anti-crossing signal in NV color centers in diamond in zero and weak magnetic fields

A. K. Dmitriev[1] and A. K. Vershovskii[1]

[1]Ioffe Institute, 194021 St. Petersburg, Russia

e-mail address: antver@mail.ioffe.ru

The response of the level anti-crossing signal to a quasi-resonant radio-frequency field, which appears in a zero magnetic field at NV color centers in diamond, is investigated. It is shown that the complex structure of this response can be explained by the Autler-Townes splitting. The possibility of controlling the parameters of the level anti-crossing signal is considered. It is shown that the slope of the central resonance recorded in this structure upon low-frequency modulation of the external magnetic field can be 2.3 times higher than the slope of the resonance recorded in the absence of an RF field. Conclusions are drawn about the nature of the level anti-crossing effect arising in zero field in NV color centers in diamond.

## I. INTRODUCTION

Negatively charged nitrogen vacancy centers, or NV centers in diamond, are a unique object that is in demand in a number of disciplines, including spintronics [1] [2], quantum informatics [3] [4], and metrology. The latter includes exact metrology of the magnetic field (MF) [5] [6], of temperature and stress in the crystal [7], as well as of frequency and time [8].

A significant number of applications of NV centers involve the use of level anti-crossing (LAC) resonances. These signals attract the attention of researchers because they can be used both in fundamental research, and in practical applications – specifically, to create sensors that do not use microwave fields to excite magnetic resonance. Most of the attention is usually focused on anti-crossing signals in strong magnetic fields (514 G – LAC in the excited state of the NV center, 1028 G – in the ground state) [1] [9] [10].

LAC also arises in zero MF [11], but its features under these conditions are much less studied. Nevertheless, these signals are of particular interest: first, due to their high symmetry (intersecting levels are identical in their properties); second, because most of the potential tasks for which sensors based on NV centers may be used (primarily biology and medicine), require them to operate in zero and weak (<1 G) MF.

The complexity of the structure of NV centers' energy levels, especially in the vicinity of LAC, allows multi-quantum and multi-frequency methods [12] of excitation of optically detected magnetic resonance (ODMR) to be used for its study. In particular, in [13], we reported the observation of high-contrast magnetically independent two-frequency resonances (more precisely, dips in conventional ODMR spectra) arising under LAC conditions in zero field upon two-frequency excitation by a microwave (MW) and radio frequency (RF) field. Similar resonances were also reported in [14]. The nature of these dips was studied in detail in [15], and in [16] a scheme of a compact frequency standard using these resonances was proposed.

On the other hand, it is known that the LAC effect can be observed not only without a microwave field, but also without applying any RF fields at all. In this sense, the LAC resonance in zero MF is no exception – when the crystal is illuminated by light resonant to the absorption band of the NV center, a slight decrease in the fluorescence level is observed in zero MF [17,18]. Due to its smallness, it makes sense to detect this signal by applying MF modulation with subsequent synchronous detection. This allows the effect of low-frequency laser intensity fluctuations (flicker noise) to be suppressed by transferring the signal to a higher-frequency region, and thus significantly increases the signal-to-noise ratio. This method was, in particular, applied in [19], and an extremely rapid drop in the signal amplitude with an increase in the modulation frequency was observed. In [19], it was concluded that the measured time constant is large even in comparison with the longitudinal relaxation time of the electron spin, and that it may be associated with the transfer of electron polarization to the nuclear spin.

It should be noted that modulation methods are also important because they are essential for constructing MF sensors. In particular, the speed of the system's response to the modulation disturbance will determine the speed of the sensor.

In [20], we investigated the change in the fluorescence spectra of NV centers in zero and weak MF under the influence of an RF field, using the amplitude modulation of the RF field. We found magnetically dependent fluorescence measurement signals, the amplitudes of which turned out to be maximum at RF field frequencies in the 4 – 6 MHz range, which in turn corresponds to the magnitude of the LAC splitting in our crystal (Fig. 1a). The range of magnetic fields in which these signals were observed also corresponded to the region of zero MF LACs $|–1,m_I\rangle$ and $|+1, m_I\rangle$, where $m_I = -1,0,+1$ is $^{14}N$ nuclear spin projection.

The LACs corresponding to different values of $m_I$ are separated by MF intervals of 0.8 G, and they presumably make similar contributions to the observed signals. For simplicity, we will further restrict ourselves to considering states $|–1,0\rangle$ and $|+1,0\rangle$ with zero nuclear spin.

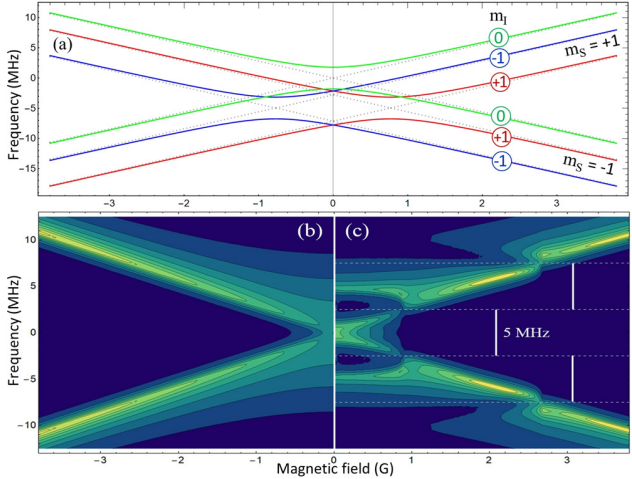

Fig.1 (a) Energy structure of the levels of the NV center in zero magnetic field in the vicinity of the intersection of levels |−1,$m_I$⟩ and |+1,$m_I$⟩. Dotted lines indicate pure states, solid lines – mixed states. The energy of pure states |±1,0⟩ in zero field is taken as zero. (b) Energy structure of levels |±1,0⟩ in the presence of transverse splitting $E$, normally distributed with a standard deviation $\sigma = (2\pi)$ 5 MHz. (c) Autler-Townes effect: disturbance of the energy structure under the influence of a radio-frequency field with a frequency of $f_{RF} = 5$ MHz (numerical simulation in accordance with Eq. (1)–(3)). The 1st and 2nd orders of perturbation are shown.

At first, the nature of these signals was unclear: as follows from considerations of symmetry, coherent superpositions of levels |−1,0⟩ and |+1,0⟩ (Fig. 1a) in zero MF should be populated equally. Therefore, it was natural to expect that the RF field resonant to these transitions cannot change the population in the system or, accordingly, change the fluorescence level.

Our further studies of optical LAC signals in zero MF under the influence of an RF field made it possible to solve this problem and develop methods for controlling the parameters of LAC signals using relatively low-frequency RF fields. This, we believe, is of interest not only for a deeper understanding of the processes occurring in NV centers and similar structures, but also for the creation of non-microwave MF sensors that are operable in zero and weak fields. The results of these studies are presented below.

## II. THEORETICAL

The level structure in the $^3A_2$ ground state is determined by the Hamiltonian $H_{tot} = H_S + H_{SI} + H_I$ (Fig.1a), where $H_S$ describes the electron spin (S = 1), $H_{SI}$ describes the hyperfine interaction with the nitrogen nucleus (I = 1 for $^{14}$N) and $H_I$ is the nuclear spin [7]. These terms can be written as follows:

$$\begin{cases} H_S = D_{gs}S_z^2 + E(S_x^2 - S_y^2) + g_S\mu_B\vec{B}\cdot\vec{S} \\ H_{SI} = A_\parallel S_z I_z + A_\perp(S_x I_x + S_y I_y) \\ H_I = PI_z^2 - g_I\mu_N\vec{B}\cdot\vec{I}, \end{cases} \quad (1)$$

where $\mu_B$ is the Bohr magneton, $\mu_N$ is the nuclear magneton, $D_{gs} = 2.87$ GHz is the longitudinal ZFS parameter, $E$ is the transverse ZFS parameter (its value depends on the parameters of a particular crystal, and, generally speaking, can be different for different NV centers in the crystal), $g_s = 2.003$ is the g-factor of the electron spin, $g_I = 0.403$ is the g-factor of the nuclear spin, $A_\parallel = -2.16$ MHz is the longitudinal hyperfine splitting constant, $A_\perp = 2.7$ MHz is the transverse hyperfine splitting constant, $P = 4.95$ MHz is the quadrupole splitting parameter. Electronic gyromagnetic ratio NV of the center $\gamma_{NV} = g_s\cdot\mu_B = (2\pi)$ 2.8 MHz/G.

The calculation of the structure of the ground state under the influence of an RF field in zero MF should take into account several effects at once. The first of these is the LAC effect itself, due to which coherent superpositions of levels |−1,0⟩ and |+1,0⟩ appear in the vicinity of the zero field. The second is the effect of Landau-Zener tunneling [21,22], due to which the prohibition on transitions between intersecting states under the influence of an RF field is lifted. The third is one of the manifestations of the dynamic Stark effect, the Autler-Townes (AT) effect, that is, the splitting of levels under the influence of a strong resonant RF field.

Earlier, H. Chen and G. Fuchs (Cornell University) calculated the effect of radio-frequency disturbances on the energy structure of the levels of the NV center in zero MF for the diamond sample used in our experiment. When considering the mechanism of action of RF disturbance on the energy structure of levels |−1,0⟩ and |+1,0⟩, the Hamiltonian of the selected states was written in the form

$$H = \begin{pmatrix} -A_\parallel + \Omega_{RF}\cos(2\pi f_{RF}t) & E \\ E & A_\parallel - \Omega_{RF}\cos(2\pi f_{RF}t) \end{pmatrix}, \quad (2)$$

where $\Omega_{RF}$ and $f_{RF}$ are the amplitude and frequency of the RF field, respectively [21,23,24]. The two levels are connected by a transverse strain field $E$. The RF field modulates the energy of the levels. The interaction probability is approximately described by the dependence [25,26]:

$$1 - e^{-\frac{E^2}{f_{RF}\Omega_{RF}}} \quad (3)$$

Multi-photon interactions of order n arise at a distance $n\cdot f_{RF}$ from the frequency of the unperturbed transition. The calculation procedure and results are described in detail in [15]. Fig. 1b and Fig 1c show our results of modeling the distribution of energy levels according to Eqs (1)-(3).

## III. EXPERIMENTAL

A detailed description of the experimental setup we used was given in [15,27]. In recent years, the installation has undergone minor modifications, and can be briefly described as follows (Fig. 2a): the radiation of a diode laser (wavelength 520 nm, power 15 mW) is collected by a collimator lens into a piece of fiber with a core diameter of 0.9 mm. The diamond crystal is glued with optically transparent glue to the other end of the segment, and is coated on the outside with a reflective dielectric coating. A $0.3 \times 0.3 \times 0.1$ mm$^3$ crystal is located in the center of the 3D system of Helmholtz rings, the currents in which are set by computer-controlled sources. The rings create the MF $B_0$, the modulus of which varies from 0 to 10 G, and the direction is set at an angle of $4\pi$. Also located around the crystal is an RF field inductor and a coil used to modulate the MF. Multiple reflections of the pumping light in the optical fiber, in the crystal, and on the reflective coating lead to its almost complete depolarization. Fluorescence radiation is collected from the same fiber. A large difference in the dimensions of the crystal and the optical fiber aperture leads to large losses of pumping light, but provides a relatively high collection efficiency of photoluminescence.

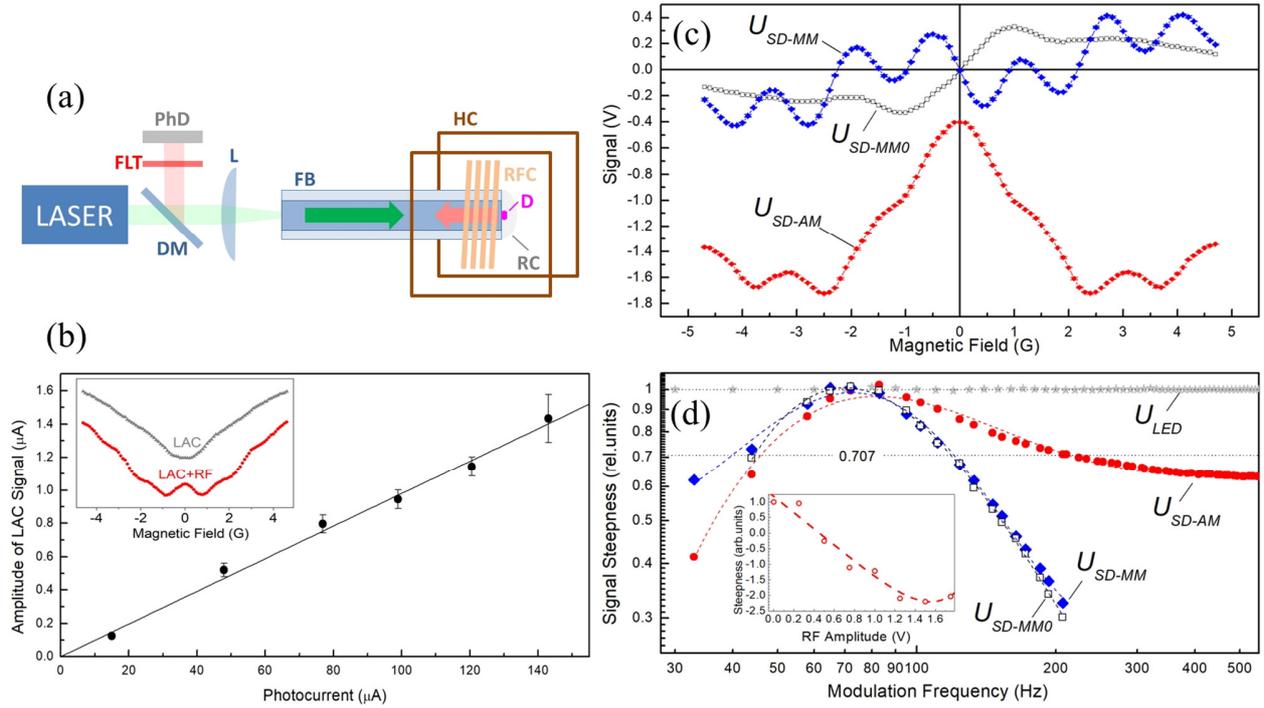

Fig. 2. (a) Simplified diagram of the experimental setup: D – diamond, RC – reflective coating, RFC – radio frequency coil, HC – Helmholtz coils (one pair of three is shown), FB – optical fiber, L – lens, DM – dichroic mirror, FLT – red optical filter, PhD – photodiode. (b) Dependence of the LAC signal amplitude on the intensity of the photocurrent at the photodetector. The inset is an example of LAC signals detected in the photocurrent (without modulation) without and with an RF field. (c) The fluorescence signal when exposed to a 5.3 MHz RF field after synchronous detection. (d) amplitude-frequency characteristic of the signal slope: $U_{SD-AM}$ – upon 100% amplitude modulation, $U_{SD-MM}$ – upon MF modulation (modulation amplitude 0.5 G). $U_{SD-MM0}$ is a signal obtained by modulating a MF without applying an RF field (LAC signal), $U_{LED}$ is a LED calibration signal. The vertical bars denote the rms determined from the scatter over ten measurements. Inset: dependence of the steepness of the $U_{SD-AM}$ signal on the amplitude of the RF field ($\Omega_{RF} \approx (2\pi)$ 2.8 MHz/V).

At the fiber output fluorescence is separated from the pumping light by a dichroic mirror. After additional filtering by a red glass filter it falls on the photodetector. As a photodetector, we use a silicon photodiode with a sensitivity of about 0.6 A/W in the red region of the spectrum. The signal from the photodetector is fed to a synchronous detector, the reference input of which is supplied with a voltage used to modulate the parameters of the RF field or MF.

The sample used in this work was provided by the Lebedev Institute of RAS. The sample preparation technology included the irradiation of SDB 1085 60/70 synthetic diamond (manufactured by Element Six) with an electron beam of intensity $5 \cdot 10^{18}$ el/cm$^2$, and subsequent annealing in for two hours in an argon atmosphere at a temperature of 800 °C. The average concentration of NV centers for such a sample should be $3\text{-}4 \cdot 10^{18}$ cm$^{-3}$. In [15], the characteristics of the sample used are given, obtained by comparing experimental data with theoretical predictions. The Gaussian distribution of the transverse splitting parameter $E$ is centered at zero with standard deviation $\sigma = (2\pi)5$ MHz (Fig. 1b).

As an RF inductor, we used two coils – a non-resonant one, characterized by a flat frequency response in the (1-7) MHz range, and a resonant one, characterized by a 50% increase in frequency response in the $5 \pm 1$ MHz range. At a frequency $f_{RF} \approx 5$ MHz, the 1 V RF field amplitude (p-t-p) at the resonant coil corresponds to $\Omega_{RF} \approx (2\pi)$ 2.8 MHz. On the non-resonant coil, 1 V corresponds to $\Omega_{RF} \approx (2\pi)1.9$ MHz.

We investigated the parameters of the fluorescence signal in the vicinity of the zero field (i.e. in the range –5 to +5 G) under the influence of the RF field, using both the amplitude modulation of the RF field and the modulation of the MF. In both cases, after amplification of the photocurrent signal, its synchronous detection was performed at the modulation frequency. Also, a direct recording of the level of fluorescence (photocurrent) was carried out.

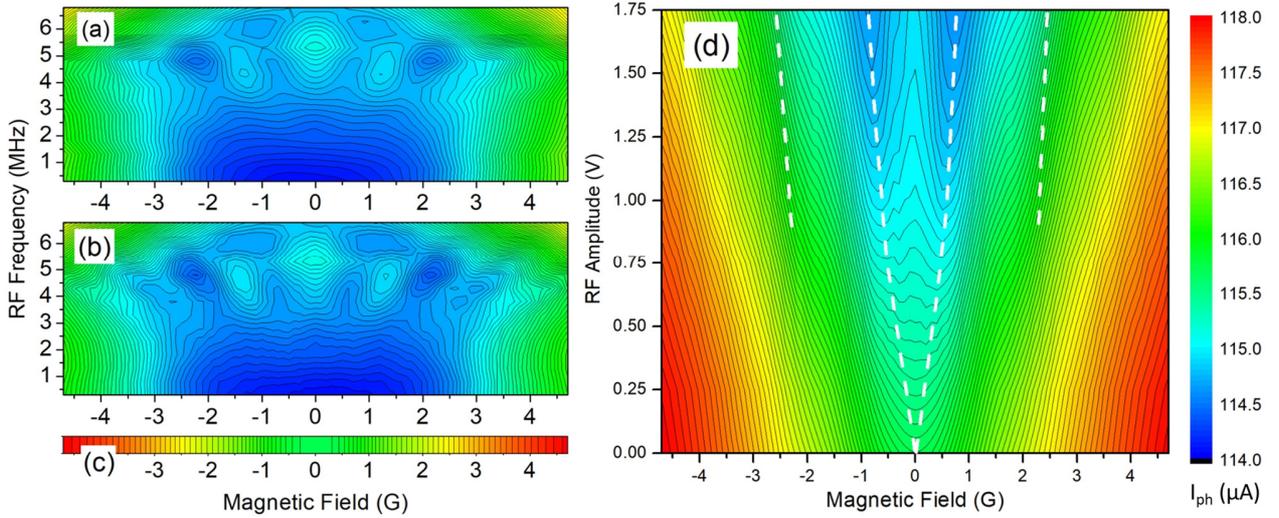

Fig. 3. Photocurrent signal upon exposure to an RF field (resonant coil) after synchronous detection: (a) upon modulation of the magnetic field: the result of numerical integration of the $U_{SD-MM}$ signal, $f_{RF}$ = 0.3–6.7 MHz, $\Omega_{RF} \approx (2\pi)$ 4.9 MHz; (b) upon modulation of the RF field amplitude: the result of adding the LAC signal to the $U_{SD-AM}$ signal, $f_{RF}$ = 0.3–6.7 MHz, $\Omega_{RF} \approx (2\pi)$ 4.9 MHz; (c) LAC signal in the absence of an RF field – the result of numerical integration of the $U_{SD-MM0}$ signal; (d) dependence on the RF field amplitude, $f_{RF}$ = 5.3 MHz, $\Omega_{RF} \approx (2\pi)$ (0–4.9) MHz. To the right of the graphs is a color scale in photocurrent units (μA).

## IV. RESULTS

A typical result of the experiment described in the previous section is illustrated in Fig. 2c, which shows signals from the output of the synchronous detector obtained with one hundred percent amplitude modulation (AM) of the RF field ($U_{SD-AM}$) and with modulation of the MP ($U_{SD-MM}$).

Fig. 2d shows the amplitude-frequency characteristics of the same signals. The main characteristic of the signal is its steepness, defined as $dI_{ph}/dB_0|_{B_0 = 0}$. The dependence of the steepness of the $U_{SD-AM}$ signal on the amplitude of the RF field is given in the inset Fig. 2b. The steepness (normalized to the steepness of the LAC signal) decreases linearly with the amplitude of the RF field, passes through zero and begins to increase in absolute value, reaching an extreme at $\Omega_{RF} \approx (2\pi)$ 4.2 MHz. In this case, the modulus of the signal steepness at the maximum exceeds that of the LAC signal by about 2.3 times.

The frequency response of the observed signals reaches its maximum value at frequencies of 70-80 Hz. At higher frequencies the amplitude of the $U_{SD-MM}$ and $U_{SD-MM0}$ signals begins to decrease proportionally to $f^{3/2}$; their frequency bandwidth is approximately 120 Hz. The amplitude of the $U_{SD-AM}$ signal, after a slight reduction, reaches the shelf and remains constant, at least up to one kilohertz. The frequency response of the electronics including the photo-amplifier, the synchronous detector, etc. was measured independently, based on the response $U_{LED}$ to the modulated light generated by the red LED. A correction has already been made to the measured dependences: this is illustrated by the $U_{LED}$ calibration signal in Fig. 2b. Thus, the low-frequency decrease in the NV centers response is not related to the hardware function of the electronics, and should be explained separately.

It can be expected that with a sufficiently small modulation amplitude, the $U_{SD-MF}$ signal will be a derivative of the $U_{SD-AM}$ signal. However, it clearly follows from Fig.2a that this is not the case: the $U_{SD-MF}$ signal has a complex quasi-periodic structure. It is logical to assume that this structure is a consequence of the presence of a signal that is not related to the RF field. This may be the LAC signal recorded in the absence of an RF field (noted as $U_{SD-MM0}$ in Fig. 2).

To test this assumption, we recorded the dependences of all the above signals on the frequency of the RF field. Measurements of the LAC signal parameters in the 50–160 mA laser current range, corresponding to the 15–143 μA range of photocurrents, showed that in this range the LAC signal width (HWHM) remains constant: $\Delta B = (2.9 \pm 0.1)$ G. The LAC amplitude, normalized to the total photocurrent intensity, is also constant: $\Delta I = (0.93 \pm 0.04)\%$. Consequently, the dependence of the signal amplitude on the intensity of the photocurrent is linear (Fig. 2b). The measurements were carried out in a magnetic field directed along the <100> axis of the crystal.

Numerical integration of the $U_{SD-MM}$ and $U_{SD-MM0}$ signals made it possible to reconstruct the dependences of the fluorescence signals on the magnetic field, while providing a significantly higher signal-to-noise ratio (compared to direct recording of the photocurrent level). The values of the photocurrent level measured at $B_0 = 0$ and averaged

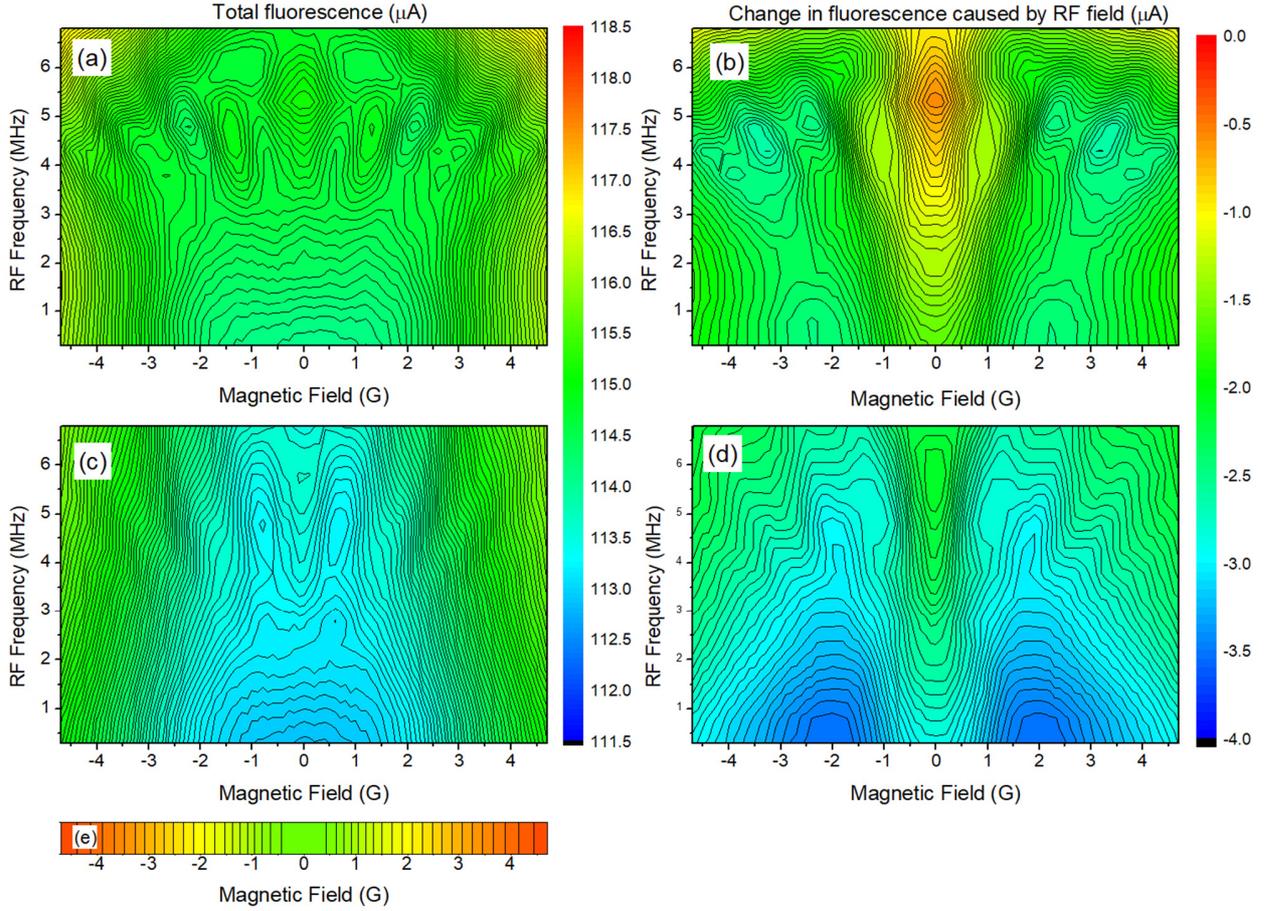

Fig. 4. Photocurrent signal (μA) when exposed to the maximum RF field amplitude ($f_{RF}$ = 0.3–6.7 MHz): (a), (c), (e) – total photoluminescence signal (the result of numerical integration of the $US_{D-MM0}$ signal); (b), (d) change in the photocurrent under the influence of the RF field; (a), (b) resonant coil, $\Omega_{RF} \approx (2\pi)$ 4.9 MHz; (c), (d) non-resonant coil, $\Omega_{RF} \approx (2\pi)$ 3.3 MHz; (e) no RF field, $\Omega_{RF}$ = 0. To the right of the graphs are color scales in photocurrent units (μA).

over a sufficiently long time were used to eliminate the uncertainty arising during integration.

Examples of such dependences obtained at the maximum amplitude of the RF field are shown in Fig. 3a-c. The graph Fig.3a is obtained by numerical integration of the $U_{SD-MM}$ signal. The chart in Fig.3b is the sum of the $U_{SD-AM}$ signal and the LAC signal (Fig.3c), which, in turn, is the result of the numerical integration of the $U_{SD-MM0}$ signal.

There is a clear agreement between the graphs Fig. 3a and Fig. 3b. Small differences are due to the smoothing arising in Fig.3a due to the non-zero amplitude of the MF modulation. Thus, the reasons for the apparent discrepancy between the signals in Fig. 2a can be considered explained, and the method used below for recovering information about the fluorescence signal by means of modulation and numerical integration is verified.

Fig. 3d shows the same signal as Fig.3b, but measured at a fixed frequency and varied amplitude of the RF field. The distance between the regions of photocurrent minima (indicated by white dashed lines) is approximately equal to $\Delta B \approx \gamma_{NV}/\Omega_{RF}$.

Fig. 4 illustrates the dependences of the total fluorescence signals obtained by numerical integration at two amplitudes of the RF field in the entire frequency range, and the signals of the change in the fluorescence under the influence of the RF field.

Since the NV center (as opposed to an isolated atom) is an object characterized by axial symmetry, it is natural to expect that the signals under study will exhibit a dependence on the direction of the MF relative to the crystal lattice. For example, [18] found a nontrivial dependence of the LAC signal shape on the direction of the polarization vector of the pump light with respect to crystal lattice and to the MF. As mentioned earlier, in our experimental configuration, we tried to achieve maximum depolarization of the pump light. Fig. 5 shows the results of measuring the photocurrent signals in three directions of the MF. During the measurements, the collinearity of the three remaining vectors was maintained – the MF vector, the MF modulation vector (in experiments where such modulation was used), and the RF magnetic component vector.

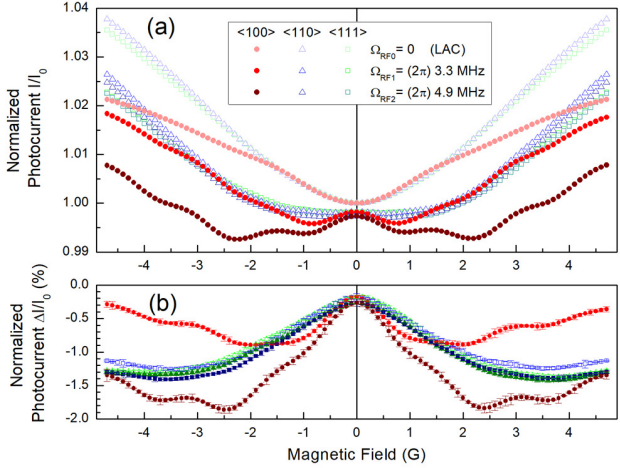

Fig. 5. Dependence of the fluorescence signal on the MF under the action of an RF field with a frequency $f_{RF}$ = 5.3 MHz and Rabi frequencies $\Omega_{RF}$ = {0, $(2\pi)$4.1 MHz, $(2\pi)$6.2 MHz}: (a) total photocurrent normalized to the value at $B_0$ = 0; (b) change in the photocurrent due to the RF field. Signals are given for three directions of the magnetic field and the collinear vector of the RF field – along the axes <100>, <110>, and <111>. The vertical bars denote the rms determined from the scatter over ten measurements.

## V. DISCUSSION

From our point of view, the most interesting result is the appearance of a narrow resonance peak (more precisely, a resonant decrease in the LAC dip) in the signal of the photoluminescence intensity under the influence of an RF field at $B_0 \approx 0$. One should also note the resonances of higher orders arising with an increase in the RF field amplitude. Their deviation from zero on the magnetic field scale, as in [15], increases nonlinearly with the frequency of the RF field. The amplitudes of these signals are maximum at RF field frequencies in the 4–6 MHz range, which corresponds to the magnitude of the LAC splitting in our crystal (Fig. 1a). The range of the MFs, in which these signals were observed, also corresponds to the region of anti-crossing of the $|\pm1, m_I\rangle$ levels.

Thus, we can conclude that all the signals we observe are due to the action of the RF field on the LAC signal recorded in the absence of the RF field. The narrow resonance peak observed in zero MF at a sufficiently large amplitude of the RF field is due to the AT splitting, and the quasi-periodic structure of the dependence of the signal on the magnetic field, observed upon modulation of the MF, is due to the higher orders of AT splitting. The AT effect splits levels in the frequency domain; nevertheless, the periodicity with respect to the MF of the structure created by AT effect corresponds to the nature of the phenomena we observe (Fig. 1b,c).

The $U_{SD-AM}$ signal obtained with the amplitude modulation technique is due only to the AT splitting. The $U_{SD-MM}$ signal, obtained with the MF modulation technique, also contains the contribution of $U_{SD-MM0}$, the LAC signal observed in the absence of an RF field.

This conclusion is qualitatively confirmed by the dependence of the waveform on the frequency and amplitude of the RF field (Fig. 3,4). It is also confirmed by the fact that the $U_{SD-AM}$ signals are characterized by a frequency dependence different from the $U_{SD-MM}$ signals (Fig. 2b). This can be explained by the fact that the $U_{SD-AM}$ signal, in contrast to $U_{SD-MM}$, is mainly due to the rapid effect of AT splitting, and is not connected (or is only partially connected) with the relatively slow process of reorientation of the electronic and nuclear moments in the structure of the NV center.

At the same time, the $U_{SD-MM}$ signal exhibits a significantly greater angular dependence than the $U_{SD-AM}$ and $U_{SD-MM0}$ signals (Fig. 5): high-contrast $U_{SD-MM}$ signals can be received only in the <100> direction. The $U_{SD-AM}$ resonance line also turns out to be the narrowest and most high-contrast in the <100> direction. This is natural, because in this configuration the projections of the magnetic field onto the NV center axis are the same for all four possible orientations of the NV centers. In other angular configurations, this condition is not met; therefore, the response line is broadened and smoothed.

Here, however, it is interesting to note that the $U_{SD-MM0}$ waveform in the range of $|\boldsymbol{B_0}|$ < 1 G is almost independent of direction. Perhaps this is due to the angular dependence of the amplitudes of anti-crossing signals between the levels $|\pm1,\pm1\rangle$, which differ in that they arise in fields of $\pm0.8$ G (Fig. 1a). The width of the observed LAC signal ((2.9 ± 0.1) G HWHM) does not allow to resolve the contributions of these anti-crossings experimentally.

From all of the above, one can draw a conclusion about the nature of the LAC effect itself in a zero field. As far as we know, there is still no definite opinion on this matter; for example, in [18] it was concluded that the "low-field feature in the magnetic spectra of NV centers in diamond" is due to "the dipole – dipole interaction between different NV centers". The authors draw this conclusion on the basis of the quadratic dependence of the effect on the concentration of NV centers discovered by them. It is also stated in [18] that "a theoretical calculation of the magnetic spectrum of an isolated NV center, for instance, by Rogers et al [28], has shown no such line".

We believe that the dependences we measured allow us to unambiguously correlate the LAC effect in zero field with anti-crossings of the levels $|-1,m_I\rangle$ and $|+1,m_I\rangle$. This is confirmed by the correspondence between the frequencies of the corresponding transitions and the frequency dependence of the response to the RF field (Fig. 1, Fig. 3).

Note that, in principle, the LAC effect in zero field can also be caused by flip-flop transitions between levels $|-1, +1\rangle$ and $|+1,-1\rangle$, accompanied by an exchange of a spin state between an electron and a nucleus. But the calculation (Fig. 1a) shows that the frequencies of such transitions at any values of $E$ do not become less than 4.32 MHz, while the frequency response of the LAC signal to RF field starts from frequencies <1 MHz.

It is clear that the LAC effect itself is not associated with real RF transitions. As follows from the linearity of the LAC signal amplitude with respect to the intensity of the photocurrent (Fig. 2b), it is not associated with nonlinear pumping processes either. Consequently, this effect is solely due to a change in the parameters of the levels $|\pm1, m_I\rangle$ (namely, their relaxation times and/or dipole moments determining the absorption of pump light) under LAC conditions. The question of exactly how the mixing of levels and their splitting due to the AT effect change the parameters of these levels still remains open.

## VI. CONCLUSIONS

The parameters of signals that arise in zero field in the fluorescence signal of NV centers of diamond when exposed to an RF field allow us to conclude that these signals are caused by the effect of Autler-Townes splitting of level anti-crossing signal. This effect allows the parameters of the anti-crossing signal to be controlled. In particular, it allows the steepness of the LAC resonance to be doubled or more, and therefore can be used to create zero and weak magnetic field sensors that do not use microwave excitation. The latter property is important for use in biomedical research, since it allows local MW heating of the tissue under study to be excluded. The peculiarities of the complex quasi-periodic signal structure due to the Autler-Townes effect of higher orders can also be used to create sensors with maximum sensitivity not at zero magnetic field, but at a selected field value in the range of several Gauss.

## ACKNOWLEDGMENTS

The reported study was funded by RFBR, project number 19-29-10004.